\documentclass[%
 reprint,
 amsmath,amssymb,
 aps,
]{revtex4-1}

\usepackage{graphicx}
\usepackage{dcolumn}
\usepackage{bm}
\usepackage{amsmath,amssymb}
\usepackage[breaklinks=true,colorlinks=true,linkcolor=blue,urlcolor=blue,citecolor=blue]{hyperref}

\begin{document}

\preprint{APS/123-QED}

\title{The influence of spin and charge fluctuations on the pressure dependence
of the N\'{e}el temperature near a quantum phase transition in rare-earth intermetallic compounds}

\author{Valery\,V. Val'kov}
\email{vvv@iph.krasn.ru}
\author{Anton\,O. Zlotnikov}
\affiliation{%
 Kirensky Institute of Physics, Russian Academy of Sciences, \\ Siberian Branch, Krasnoyarsk, 660036 Russia
}%


\begin{abstract}
In the one-loop approximation for the periodic Anderson model the contributions of spin and charge fluctuations to
the renormalization of the antiferromagnetic order parameter are calculated. It is shown that taking into account
the fluctuation corrections allows to quantitatively describe the pressure dependence of the N\'{e}el temperature
observed in quasi-two-dimensional intermetallic antiferromagnet with heavy fermions CeRhIn$_5$.
\begin{description}
\item[PACS numbers]
\verb+71.27.+a+, \verb+75.30.Mb+, \verb+74.40.Kb+.
\end{description}
\end{abstract}

\pacs{Valid PACS appear here}

\maketitle

\section{Introduction}

The considerable interest in the properties of heavy-fermion antiferromagnets is due to their unconventional superconductivity, quantum phase transitions, pronounced competition between a tendency to magnetic ordering and Kondo fluctuations, and coexistence of superconductivity and antiferromagnetism. The quantum phase transitions are initiated by an external or chemical pressure and are accompanied by the variations in the ground state structure, which leads to the change in the characteristics of materials. In particular, in heavy-fermion CeCu$_{6-x}$Au$_x$ and YbRh$_2$Si$_2$ metals, a passage through the quantum critical point is accompanied by destruction of the long-range antiferromagnetic~(AFM) order varying the control parameters, specifically, dopant concentration~$x$ and magnetic field~\cite{Gegenwart-08}. Under pressure, the phase diagrams of the Ce-based compounds, including CePd$_2$Si$_2$, CeIn$_3$~\cite{Mathur-98}, CeRhIn$_5$~\cite{Hegger-00,Kohori-00}, and CePt$_2$In$_7$~\cite{Bauer-10}, contain a superconductivity dome in the vicinity of the expected quantum critical point.

The nature of magnetic ordering is one of the most important problems of physics of heavy-fermion systems. If we assume that the long-range AFM order is initiated by the RKKI indirect exchange interaction and Kondo fluctuations tend to destruct the magnetic ordering, then at such a scenario of the competition~\cite{Doniach-77} at the quantum critical point, simultaneously with the destruction (occurrence) of antiferromagnetism, the Kondo regime can be established (suppressed)~\cite{Coleman-01,Si-01} and the transition from localized to delocalized electrons can occur~\cite{Kuramoto-13}.

According to the modern concepts, 4f electrons in the AFM phase of cerium compounds are quasi-localized and form the coherent heavy-fermion state. This was confirmed by the experimental data, which show that even in the CeRhIn$_5$ AFM phase the effective and cyclotron electron masses are larger than the free electron mass~\cite{Shishido-05,Knebel-08}. According to study~\cite{Miyake-14}, the mixed-valence regime can be implemented in this compound. Therefore, it is reasonable to investigate the formation of magnetic ordering using the periodic Anderson model (PAM) in the regime when the localized bare level is close to the Fermi level. Such an approach allows describing the strong renormalization of electron mass and the Fermi surface topology variation at a quantum critical point without using the Kondo breakdown scenario~\cite{HWatanabe-08,SWatanabe-10,Val'kov-13}.

The occurrence of the AFM phase in the PAM was demonstrated, e.g., in the Hatree-Fock approximation~\cite{Leder-78} and using the slave-boson technique~\cite{Moller-93,Spalek-98}. It should be noted that in the considered approaches N\'{e}el temperature $T_{\text{N}}$ takes small values only around the quantum phase transitions (see also~\cite{Sacramento-03}). It means that even minor variations in the external factors, e.g., pressure, can induce the quantum phase transition from the AFM to paramagnetic (PM) phase. However, in many heavy-fermion compounds with $T_{\text{N}}$ of no more than few Kelvin degrees, antiferromagnetism is sufficiently stable against pressure variations.

The alternative approach to finding a magnetic instability point is the calculation of dynamic magnetic susceptibility. In the PAM, the dynamic magnetic susceptibility was calculated using equations of motion for irreducible Green's functions~\cite{Foglio-78} and in the random phase approximation~\cite{Mills-80}. In study~\cite{Kuramoto-81}, it was proposed to calculate the dynamic magnetic susceptibility using the perturbation theory for the hybridization interaction. It was demonstrated that in the mixed-valence regime the effective interaction caused by hybridization between localized and itinerant electrons suppresses any magnetic fluctuations. In the limit $U \to \infty$, where $U$ is the parameter of on-site Coulomb interaction, the method for calculating the dynamic magnetic susceptibility was developed on the basis of a diagram technique for Hubbard operators within the Hubbard model and $t-J$ model~\cite{Izyumov-90}. In study~\cite{Val'kov-10}, this method was applied to determine the dynamic magnetic susceptibility in the PAM paramagnetic phase.

\begin{figure}[htb!]\center
\resizebox{0.2\textwidth}{!}{%
  \includegraphics{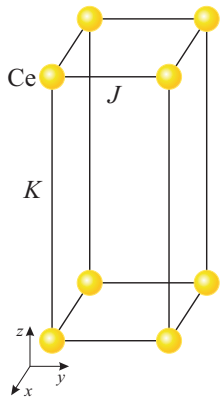}
}
\caption{\label{quasi2D} Schematic of the quasi-two-dimensional structure of Ce115 compounds. $J$ and $K$ are the constants of magnetic exchange between Ce ions.}
\end{figure}

Heavy-fermion compounds, e.g., Ce$_n$T$_m$In$_{3n+2m}$ \cite{Thompson-12}, have a quasi-two-dimensional (quasi-2D) structure schematically illustrated in Fig.~\ref{quasi2D}. As is known, the N\'{e}el temperature of a quasi-2D Heisenberg antiferromagnet is approximately determined as $T_\text{N} = \pi J/\left( \ln(J/K) + c \right)$, where $J$ is the parameter of exchange between the nearest ions in the (xy) plane, $K$ is the value of exchange interaction between the nearest neighbors along the z axis, and $c$ is the constant depending on the lattice type~\cite{Val'kov-11}. This formula indicates a decrease in the transition temperature relative to the isotropic case. In study~\cite{Das-14}, using the neutron spectroscopy data and Heisenberg model, the parameters of exchange between Ce ions in the CeRhIn$_5$ compound were found to be $J = 0.74$ meV and $K = 0.1$ meV.

It should be noted that in the most interesting case of strong electron correlations for the electron states of rare-earth ions and the one-electron excitation energy close to the Fermi level (the mixed-valence case), it is convenient to divide the hybridization processes into  high- and low-energy ones~\cite{Val'kov-08}. The high-energy processes are the transitions at which, due to the strong correlations, the energy of a system changes by a value much larger than the hybridization interaction parameter. This great energy difference makes it possible to take into account the above-mentioned hybridization mixing using the operator perturbation theory in the atomic representation and to obtain the effective Hamiltonian describing, in particular, the exchange coupling between quasi-localized states of rare-earth ions. The parameter of this interaction is determined as $J \sim V^4/U^3$, where  $V$ is the intensity of hybridization between localized and itinerant electrons. The rest low-energy contributions determine the properties of the mixed-valence regime. Note that in the Ce115 compounds (e.g., CeRhIn$_5$), the role of localized electrons is played by Ce 4f electrons and the collective states are formed mainly by In p electrons~\cite{Elgazzar-04}.

In this work, using the effective PAM, which explicitly takes into account the exchange interaction between 4f electrons, we obtain a pressure dependence of the N\'{e}el temperature for quasi-2D cerium intermetallic compounds. This dependence is not only qualitatively consistent with the data reported in~\cite{Park-09}, but also describes well the experimental results. The pressure dependence of the N\'{e}el temperature consists of two portions. The first portion is characterized by a linear decrease in the N\'{e}el temperature with increasing pressure. Such a behavior is qualitatively reproduced with disregard of hybridization. The second portion shows a sharp break of antiferromagnetism. We show that this dependence is formed only with regard to the low-energy hybridization processes. In this case, two channels affecting the magnetic ordering occur: the exchange interaction of 4f electrons tends to establish the AFM ordering, whereas the low-energy hybridization of f and p electrons can both promote and suppress antiferromagnetism. In this study, we estimate partial contributions of these microscopic mechanisms to the experimental effective parameters of exchange interactions in the CeRhIn$_5$ compound~\cite{Das-14}.

\section{Model and method}

The Hamiltonian of the effective PAM, which takes into account the exchange interaction between 4f electrons for a simple cubic lattice, is
\begin{eqnarray}
\label{HamiltonianPAMeff}
&& \widehat{H}_{\text{eff}} = \sum_{k\sigma} \xi_{k}
c_{k\sigma}^{\dag}c_{k\sigma} +
\sum_{m\sigma} \xi_{L \sigma} X_{m}^{\sigma \sigma} +
\nonumber \\
 & + & \frac{1}{2} \sum_{m \ne l} J_{ml} \left( \vec{S}_{m} \vec{S}_{l} -
\frac{1}{4}\hat{N}_{m}\hat{N}_{l} \right) +
\nonumber \\
& + & \frac{1}{\sqrt{N}}\sum_{km\sigma} \left[ e^{-i\vec{k}\vec{R}_m} V_{k}
c_{k\sigma}^{\dag}X_{m}^{0\sigma}
+ h.c. \right].
\end{eqnarray}
The first term in the Hamiltonian describes a subsystem of itinerant electrons (In p electrons) in the quasi-momentum space with bare energy $\xi_{k}$ counted from chemical potential $\mu$. Localized 4f electrons corresponding to Wannier cell~$m$ are described in the atomic representation using the Hubbard operators $X_{m}^{n s}=|n; m \rangle \langle m; s|$, where $|n; m \rangle$ is the atomic states without f electron ($|0; m \rangle$) and with one f electron ($|\sigma; m \rangle$) and with different spin moment projections $\sigma$. From the doublet in the crystal field, the f level with $j=5/2$ is taken into account. For the bare energy of 4f electron $\xi_{L \sigma}$, the self-consistent field is introduced. The exchange interaction is specified by parameter $J_{ml}$, $\vec{S}_{m}$ is the quasi-spin vector operator of f electron, and $\hat{N}_m$ is the operator of the number of localized electrons on site~$m$. The fourth term in the Hamiltonian determines the low-energy hybridization processes between localized and itinerant electrons with intensity $V_{k}$; $N$ is the number of sites in the lattice.

\begin{figure}[htb!]\center
\resizebox{0.4\textwidth}{!}{%
  \includegraphics{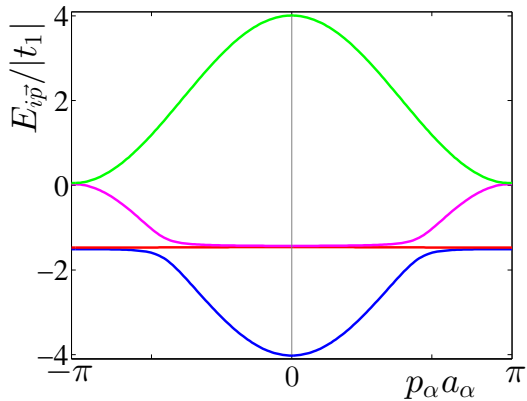}
}
\caption{
General view of quasiparticle spectrum in the antiferromagnetic phase of the periodic Anderson model with regard to hybridization and hoppings only in the $(xy)$ plane. \label{fspectr}}
\end{figure}

To describe the magnetic properties, we apply a diagram technique for the Hubbard operators~\cite{Zaitsev-07,Val'kov-04}, which will be used for calculating the Matsubara Green's function in the atomic representation~\cite{Zaitsev-75}
\begin{equation}
D_{\perp}\left(m \tau; m' \tau'\right) = - \left\langle T_{\tau} X_m^{\uparrow\downarrow}\left(\tau\right) X_{m'}^{\downarrow \uparrow}\left(\tau'\right) S(\beta)  \right\rangle_{0, c}.
\end{equation}
The time-dependent Hubbard operators are taken in the interaction representation;  $T_{\tau}$ is the time-ordering operator. Index $0$ indicates that averaging is made with regard to the Hamiltonian that describes noninteracting itinerant and localized electrons. Only connected diagrams are taken into account. The scattering matrix has the form
$S(\beta) = T_{\tau} \exp \left( -\int_0^{\beta} \widehat{H}_{int}(\tau) d\tau \right)$, where interaction operator $\widehat{H}_{int}$ involves the operators of hybridization and exchange interactions and $\beta = 1/T$ is the inverse temperature.

As is known~\cite{Zaitsev-76}, the Fourier image of Green's function can be presented in the form $D_{\perp}\left(q\right) = G_{\perp}\left(q\right)P\left(q\right)$, where $q~=~(\vec{q}, i \omega_m)$, $i\omega_m$ are the even Matsubara frequencies and $P\left(q\right)$ is the force operator. Then, the Dyson equation for the function $G_{\perp}\left(q\right)$ is
\begin{eqnarray}
\label{Eq_Dyson}
G_{\perp}\left(q\right) & = & G^{(0)}_{\perp}\left(q\right) + G^{(0)}_{\perp}\left(q\right) \Sigma\left(q\right) G_{\perp}\left(q\right),
\\
\label{Eq_Larkin}
G^{(0)}_{\perp}\left(q\right) & = & G^{(0)}\left(i\omega_m\right) + G^{(0)}\left(i\omega_m\right) P\left(q\right) J_{\vec{q}} G^{(0)}_{\perp}\left(q\right),
\end{eqnarray}
where $G^{(0)}\left(i\omega_m\right) = (i\omega_m-2\tilde{h})^{-1}$ is the bare Green's function with the self-consistent exchange field $\tilde{h}$, $J_{\vec{q}}$ is the Fourier image of the exchange integral, and all the corrections related to hybridization are contained in  mass operator $\Sigma\left(q\right)$ and force operator $P\left(q\right)$. For convenience, below we explicitly distinguish the bare vertex and corrections related to the hybridization interaction $P\left(q\right) = \left\langle S_m^z \right\rangle + \delta P\left(q\right)$ in the force operator.

The solution of Eq.~\eqref{Eq_Dyson} is given by
\begin{eqnarray}
\label{Ans_G_per}
G_{\perp}\left(q\right) =   \frac{G^{(0)}\left(i\omega_m\right)}{1-\left[\Sigma\left(q\right)+J_{\vec{q}}P\left(q\right)\right]G^{(0)}\left(i\omega_m\right)}.
\end{eqnarray}
In the limit $V~\to~0$ the solution acquires a simple form corresponding to the Tyablikov approximation for the Heisenberg model. In the steady-state case, this function can have the features related to the formation of ferromagnetic or AFM ordering at the instability point. The exchange interaction between 4f electrons, which is induced by high-energy hybridization processes, leads to antiferromagnetism and the low-energy hybridization processes affect the initial AFM state. In this case, mass operator $\Sigma\left(q\right)$ indicates the occurrence of a new effective exchange due to the residual part of the hybridization interaction and corrections $\delta P\left(q\right)$ renormalize the bare vertex.

The quasi-2D structure of the Ce115 cerium compounds is schematically shown in Fig. 1. Parameter $J$ of the exchange between the nearest Ce ions in the ($xy$) plane is significantly larger than the analogous parameter $K$ along the $z$ axis. In study~\cite{Das-14}, for CeRhIn$_5$ the weak exchange between next-to-nearest neighbors along the $z$ axis was also taken into account, which allowed describing the incommensurate magnetic structure at ambient pressure. In this study, we limit the consideration to the account for the exchange interaction only between the nearest neighbors, since under pressure the CeRhIn$_5$ AFM structure becomes commensurate~\cite{Yashima-09}. Then, the long-range exchange parameters can be ignored. It should be noted that the low-energy mixing of f and p electrons, as well as and hoppings of itinerant In electrons, are limited only by the ($xy$) plane.

To describe antiferromagnetism, it is convenient to pass to the two-sublattice representation. Then, the bare energy of localized electrons in the F sublattice, the magnetization of which is codirectional to the $z$ axis, is determined as $\xi_{F\sigma} = E_0 - \mu - (2J+K)(n_{\text{L}}/2+\eta_{\sigma}R)$. Here, $n_{\text{L}}$ is the f-electron concentration and $R = \langle S_f^z \rangle$, $\eta_{\sigma} = 1 (-1)$ at $\sigma = \uparrow (\downarrow)$. The bare energy of f electrons in the G sublattice is $\xi_{G\sigma} = \xi_{F\bar{\sigma}}$ where $\bar{\sigma}$ denotes the opposite direction of $\sigma$. In the description of a quasi-2D antiferromagnet, all the quantities in Dyson equations (\ref{Eq_Dyson}, \ref{Eq_Larkin}) should be replaced by matrices. Then, the matrix Green's function is determined as
\begin{eqnarray}
\label{MatrixG}
\widehat{G}_{\perp} =
\left( \begin{matrix}
\widehat{G}_{\perp}^{FF}&
\widehat{G}_{\perp}^{FG}
\vspace{6pt} \\
\widehat{G}_{\perp}^{GF}&\widehat{G}_{\perp}^{GG}
\end{matrix} \right),
\, \, \hfill
\widehat{G}_{\perp}^{AB} =
\left( \begin{matrix}
G_{\perp}^{A_1B_1}&
G_{\perp}^{A_1B_2}
\vspace{6pt} \\
G_{\perp}^{A_2B_1}&G_{\perp}^{A_2B_2}
\end{matrix} \right),
\end{eqnarray}
where $A, \, B = F, \, G$. Notations $F_n$ and $G_n$ ($n = 1, 2$) indicate that this Green's function is built on the operators belonging to the F or G sublattice and corresponding to the $n$ plane of the unit cell presented in Fig. 1. The matrices for the mass and force operator components are written in a similar way.
Taking into account the exchange parameters denoted on Fig. \ref{quasi2D} an interaction matrix is given by:
\begin{eqnarray}
\label{MatrixJ}
\widehat{J} =
\left(
\begin{matrix}
\widehat{O}&
\widehat{I}
\\
\widehat{I}&\widehat{O}
\end{matrix} \right),
\, \, \, \, \, \,  \, \, \, \hfill
\widehat{I} =
\left( \begin{matrix}
J_{\vec{q}}&
K_{\vec{q}}
\vspace{4pt} \\
K_{\vec{q}}&J_{\vec{q}}
\end{matrix}\right),
\end{eqnarray}
where $\widehat{O}$ is a null matrix.

\begin{figure}[htb!]
\begin{center}
\resizebox{0.48\textwidth}{!}{%
  \includegraphics{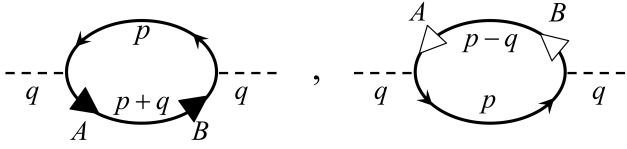}
}
\caption{Diagrams for component $AB$ of the mass operator.}
\label{Diag_S}
\end{center}
\end{figure}

\begin{figure}[htb!]
\begin{center}
\resizebox{0.48\textwidth}{!}{%
  \includegraphics{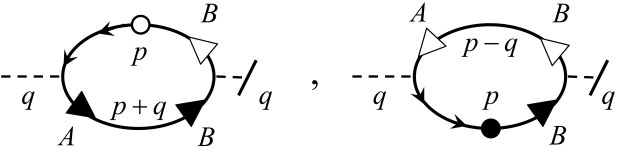}
}
\caption{Diagrams for component $AB$ of the force operator.}
\label{Diag_P}
\end{center}
\end{figure}

Figures \ref{Diag_S} and \ref{Diag_P} show the general view of diagrams for arbitrary components $A_nB_n$ of the matrix mass and force operators. Note that the expressions are independent of number~$n$ of the plane in the quasi-2D unit cell, so below this index will be omitted. In the figures, solid lines with two arrows $\vartriangleright$ and $\blacktriangleright$ indicate propagators in the Hubbard-I approximation for localized electrons with spin moment projections $\uparrow$ and $\downarrow$, respectively, with regard to hybridization. The bare Green's functions for f electrons are indicated by the solid line with one arrow $\vartriangleright$ ($\blacktriangleright$). The solid line with two arrows \textbf{$\succ$} indicates any of the four bold propagators for itinerant electrons in the two-sublattice representation. Symbols $\circ$ and $\bullet$ indicate the Hubbard vertex factors $F_{0\sigma} = \left\langle X_m^{00} + X_m^{\sigma \sigma} \right\rangle_0$ for the corresponding electron spin directions. The summation is made over the internal momenta $p = (\vec{p}, i\omega_n)$, where $\omega_n$ are the odd frequencies. The total number of diagrams for the matrix mass and force operators is 64.

The analytical expression for mass operator component $FF$ obtained from the diagrams in Fig. \ref{Diag_S} is
\begin{align}
& \Sigma^{FF}(q) = - \frac{T}{2N} \sum_{p \sigma} \eta_{\sigma} G^{FF}_{\bar{\sigma}}(p+\eta_{\sigma}q)  \times
\nonumber \\
& \times \left[ \left(V_{\vec{p}}+W_{\vec{p}}\right)^2 G^{\alpha \alpha}_{\sigma}(p) + \left(V_{\vec{p}}-W_{\vec{p}}\right)^2 G^{\beta \beta}_{\sigma}(p) + \right.
\nonumber \\
& + \left.  \left(W_{\vec{p}}^2-V_{\vec{p}}^2\right) \left(G^{\alpha \beta}_{\sigma}(p) + G^{\beta \alpha}_{\sigma}(p) \right) \right].
\end{align}
The functions $G^{AB}_{\sigma} \left(p\right)$ and $G_{\sigma}^{\nu\mu} \left(p\right)$ ($A, \, B = F, \, G$; $\nu, \, \mu = \alpha, \, \beta$) are the propagators for localized and itinerant electrons in the Hubbard-I approximation in the two-sublattice description, respectively, and $V_p$ and $W_p$ are the Fourier images of hybridization integrals inside a sublattice and between sublattices, respectively.

Using the matrix Dyson equations, it is easy to obtaine the Green's function $D^{F_1F_1}_{\perp}$ the poles of which determine the spin-wave excitation spectrum for a quasi-2D antiferromagnet with regard to hybridization processes between localized and itinerant electrons. These poles are described well by approximate analytical expressions:
\begin{equation}
\omega_{1,2 \, \vec{q}} = \omega_{01,2}\left(\vec{q}\right) + \delta\omega_{1,2 \, \vec{q}},
\end{equation}
where $\omega_{01,2}\left(\vec{q}\right)$ are the bare branches of the spectrum of a quasi-2D Heisenberg antiferromagnet (without hybridization) and the corrections
\begin{align}
\label{domega}
&\delta\omega_{i \, \vec{q}} = d_{i \, FF}+d_{i \, GG} \pm \left( K_{i \, FG} + K_{i \, GF} \right) - \frac{1}{\omega_{0i}\left(\vec{q}\right)} \times
\nonumber \\
&\times \left\{ \left(4J+2K\right)R\left[ d_{i \, GG} - d_{i \, FF} \pm \left( K_{i \, FG} + K_{i \, GF} \right) \right] + \right.
\nonumber \\
&\left. + (J_{\vec{q}}+K_{\vec{q}})R\left[ d_{i \, FG} - d_{i \, GF} \pm \left( K_{i \, FF} + K_{i \, GG} \right) \right] \right\}
\end{align}
describe the effect of hybridization interaction on the magnon spectrum. We introduced the designations
\begin{eqnarray}
d_{i \, AB} & = & \Sigma^{AB}\left(\omega_{0i}\left(\vec{q}\right)\right)/2 +
J_{\vec{q}} \, \delta P^{A\bar{B}}\left(\omega_{0i}\left(\vec{q}\right)\right)/2,
\\
 K_{i \, AB} & = & K_{\vec{q}} \, \delta P^{AB}\left(\omega_{0i}\left(\vec{q}\right)\right)/2,
\end{eqnarray}
 where the mass and force operator components are calculated after the analytical continuation and $\bar{F} = G$, $\bar{G} = F$.

The antiferromagnetic order parameter is defined as $R = n_{\text{L}}/2-\langle X_{f_1}^{\downarrow\downarrow} \rangle$, where
\begin{equation}
\left\langle X_{f_1}^{\downarrow \downarrow} \right\rangle = -\frac{T}{N/2} \sum_{q} e^{-i \omega_n \delta}  D^{F_1F_1}_{\perp} \left(q\right), \, \, \, \delta \to 0.
\end{equation}
Using the obtained self-consistent equation, the N\'{e}el temperature in the limit $R \to 0$ is determined.

\section{Results and discussions}

\subsection{Spin-wave spectrum of quasi-two-dimensional heavy-fermion antiferromagnets}

It follows from the obtained expressions for the spin-wave spectrum that the Goldstone theorem  about the existence of gapless magnon excitations in the AFM phase at $U \to \infty$ with regard to hybridization of f and p electrons is valid. The occurrence of two branches in the magnon spectrum is caused by different characters of rotation of spin moments on the nearest sites  in the $(xy)$ plane and along the $z$ axis with regard to the quasi-two-dimensionality in the antiferromagnet (inphase and antiphase, respectively). The branch $\omega_{1 \, \vec{q}}$ indicated by the superscript in formula ~\eqref{domega} is Goldstone. For the branch $\omega_{2 \, \vec{q}}$, the excitations are separated by an energy gap.

\begin{figure}[htb!]\center
\resizebox{0.39\textwidth}{!}{%
  \includegraphics{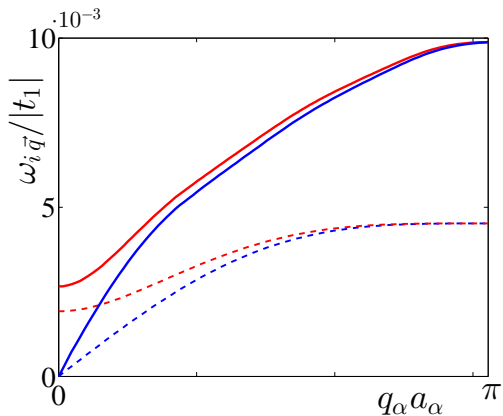}
}
\caption{Spin-wave spectrum of the quasi-two-dimensional structure with regard to hybridization between localized and itinerant electrons (solid lines) and bare spectrum of localized electrons (dashed lines) for the (111) principal direction of the antiferromagnetic Brillouin zone. Concentration of 4f electrons is $n_{\text{L}} \approx 0.7$ (see text for details).
\vspace{0.0cm}
}\label{mspectr_nL07}
\end{figure}

We consider the case when the Fermi level lies close to~$E_0$ and crosses the weak-dispersion region in Fig. \ref{fspectr}, where the general structure of the quasi-particle spectrum in the PAM is shown. Then, heavy fermions can be formed in the magnetically ordered phase the mass of which exceeds the mass of free electrons. We suggest that this state is implemented in CeRhIn$_5$ near atmospheric pressure.

The spin-wave spectrum for the principle direction of the AFM Brillouin zone, which corresponds to the f-electron concentration $n_{\text{L}} \approx 0.7$, is shown in Fig.~\ref{mspectr_nL07}. In the figure, $q_{\alpha}$ is the wave vector component and $\alpha = x, \, y, \, z$, $a_{\alpha}$ is one of the unit cell parameters. Dashed lines indicate magnon energies $\omega_{0i}\left(\vec{q}\right)$ for the quasi-2D structure with disregard of the hybridization between p and f electrons. Solid lines are the branches $\omega_{i\vec{q}}$ of the spin-wave spectrum with regard to the hybridization interaction. It can be seen that the low-energy hybridization processes lead to a significant increase in the spin-wave stiffness $\kappa$ for the Goldstone mode $\omega_{1q} = \kappa q$ (at small $q$ values) and magnon energy.

\begin{figure}[htb!]\center
\resizebox{0.39\textwidth}{!}{%
  \includegraphics{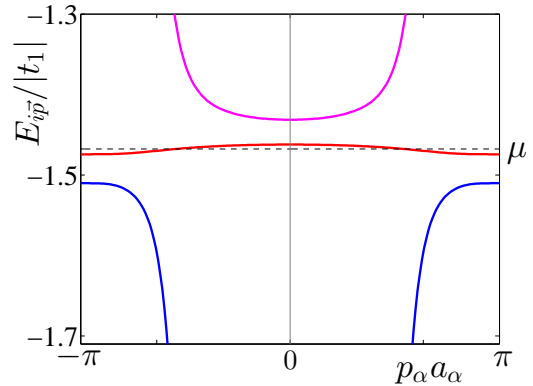}
}
\vspace{0.1cm}
\caption{
Fermi spectrum for the f-electron concentration $n_{\text{L}} \approx 0.7$.  Filling level $\mu$ is indicated by the dashed line. \label{fspectr_nL07}
\vspace{0.0cm}}
\end{figure}

The interaction parameters for $n_{\text{L}} \approx 0.7$ were chosen in the form $V = 0.3 |t_1|$, $J = 0.004 |t_1|$, $K = J/10$, where $t_1$ is the matrix element of hoppings of itinerant electrons between the nearest sites ($t_1<0$), $E_0 = 1.5t_1$ is the energy of the f-level, and $n_{\text{e}} = 1.2$ is the total electron concentration. Comparison of the model law of dispersion of itinerant electrons~$\xi_k$ and the dispersion dependencies for In p electrons in CeRhIn$_5$ obtained using ab initio calculations yields $|t_1| \approx 0.1-0.3$ eV \cite{Elgazzar-04}. The chosen parameters correspond to the Fermi excitation spectrum presented in Fig.~\ref{fspectr_nL07}.

It should be noted that in the regime when the localized electron subsystem is almost completely filled ($n_{\text{L}}\approx1$) the magnon spectrum almost does not change with regard to the hybridization interaction.

Modification of the spin-wave spectrum is explained by the occurrence of the additional effective exchange interaction due to low-energy hybridization processes. Indeed, the comparison of Dyson equation~\eqref{Eq_Dyson}, which contains the mass operator, and equation~\eqref{Eq_Larkin}, which takes into account the exchange interaction between 4f electrons and vertex factors, shows that the mass operator components work as the effective exchange interaction. The exchange parameters of the effective interaction between different lattice sites can be estimated as
\begin{eqnarray}
\label{Eff_Exch_par}
A_{f,g} = \frac{1}{N/2} \sum_{\vec{q}} e^{i \vec{q} (\vec{R}_f-\vec{R}_{g})} \Sigma^{FG} (\vec{q}),
\\
I_{f,f'} = \frac{1}{N/2} \sum_{\vec{q}} e^{i \vec{q} (\vec{R}_f-\vec{R}_{f'})} \Sigma^{FF} (\vec{q}),
\end{eqnarray}
where sites $f$ and $f'$ belong to the F sublattice and sites $g$, to the G sublattice. It should be noted that the effective interaction occurs only between ions lying in the $(xy)$ plane, since the low-energy hybridization processes are limited by this plane. Thus, the exchange interaction along the $z$ axis with parameter $K$ determines the quasi-2D character of the systems under study and exchange parameter $J$ in the $(xy)$ plane is renormalized due to the hybridization interaction. In this case, the exchange between the next-to-nearest neighbors is also formed by the expense of the effective interaction in the $(xy)$ plane.

The effect of hybridization between p and f electrons on the characteristics of the AFM phase, including the spin-wave stiffness, AFM order parameter, and N\'{e}el temperature, is determined by the signs of parameters $A_{f,g}$ and $I_{f,f'}$. At $A_{f,g}~>~0$ and $I_{f,f'}<0$, the hybridization processes promote the AFM ordering and the spin-wave spectrum acquires the form presented in Fig.~\ref{mspectr_nL07}. If the magnon energy decreases with regard to hybridization, as it happens near the quantum critical point (see the next paragraph), then the AFM exchange between sublattices in the $(xy)$ plane weakens, since $A_{f,g}<0$, and the exchange inside the sublattice satisfies the inequality $I_{f,f'}>0$. Thus, it is important that the introduced effective interaction parameters depend on the localized level position, density of states, and temperature.

\subsection{Pressure dependence of the N\'{e}el temperature}

The low-energy hybridization processes not only lead to the quantitative variation in the parameters of the AFM phase, but also qualitatively change the behavior of these parameters. Let us consider the dependence of N\'{e}el temperature $T_{\text{N}}$ on pressure $P$ in heavy-fermion quasi-2D Ce-based antiferromagnets, such as CeRhIn$_5$. According to the experimental data, the N\'{e}el temperature linearly decreases with increasing pressure in a fairly wide pressure range~\cite{Park-09}. At the critical pressure, the N\'{e}el temperature turns to zero and the long-range AFM order is destroyed. Note that the N\'{e}el temperature in these materials is no higher than few Kelvin degrees.

It is assumed that the pressure growth leads to an increase in energy $E_0$ of a 4f electron on the positively charged Ce ion due to enhancing Coulomb interaction with the negatively charged environment. Since this interaction, including that between sites, is the strongest in these systems, the effect of the growth of $E_0$ prevails over the growth of hybridization intensity and hoppings with increasing pressure.

\begin{figure}[htb!]\center
\resizebox{0.45
\textwidth}{!}{%
  \includegraphics{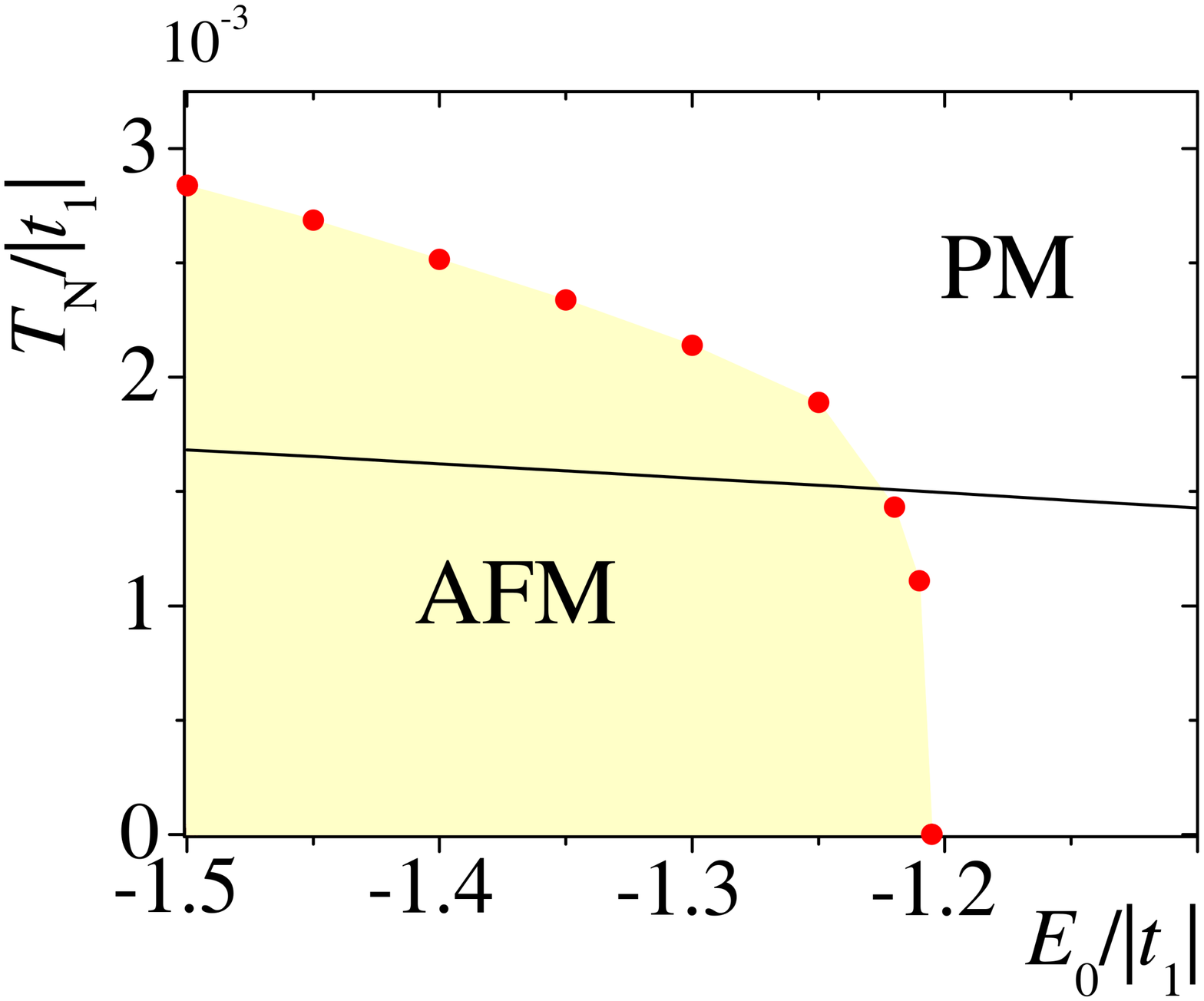}
}
\caption{Dependence of the N\'{e}el temperature on the energy of 4f electrons (pressure) with regard to hybridization (dots) and with disregard of it (solid line). The shaded area indicates the antiferromagnetic (AFM) phase and the white area, the paramagnetic (PM) phase. \label{fig_TN_E0}}
\end{figure}

Figure \ref{fig_TN_E0} shows the dependence of the N\'{e}el temperature on the bare energy of 4f electron (pressure). Dots indicate the dependence with regard to the low-energy hybridization processes between f and p electrons. This curve separates the regions of implementation of the AFM phase (dashed area) and the PM phase. The solid line shows the dependence $T_{\text{N}}(P)$ with disregard of the hybridization interaction. It can be seen that in this case the N\'{e}el temperature linearly decreases with increasing pressure. Such a behavior is related to a linear decrease in 4f-electron concentration. The renormalized curve also contains a linear portion; however, the N\'{e}el temperature grows due to hybridization. It is more important that the account for the hybridization interaction leads to destruction of antiferromagnetism with increasing pressure and the dependence $T_{\text{N}}(P)$ becomes consistent with the experiment. The quantitative consistency of the results with the data for CeRhIn$_5$ is reached if we take $|t_1|~\approx~0.14$~eV. This estimation is adequate for heavy-fermion systems. It should be noted that at the quantum phase transition point in pressure from the AFM to PM phase the Fermi surface broadens and the effective electron mass strongly grows~\cite{Val'kov-13}, as was experimentally observed in~\cite{Shishido-05}.

At $E_0 = -1.5|t_1|$, the spin-wave and Fermi spectra are presented in Figs. \ref{mspectr_nL07} and \ref{fspectr_nL07}, respectively. The localized electron concentration is $n_{\text{L}} \approx 0.7$. It can be seen in Fig. \ref{fig_TN_E0} that this point corresponds to the pressure similar to atmospheric. Using formula \eqref{Eff_Exch_par}, we estimate the effective parameter of exchange between the nearest ions caused only by the low-energy hybridization processes: $\tilde{A}_1 = 0.0036|t_1|$, which corresponds to the AFM exchange. When estimating this parameter, we excluded the coefficient that takes into account the vertex factor. Then, the total value of exchange interaction between the nearest ions in the $(xy)$ plane  is written in the form $J_{\text{eff}} = J + \tilde{A}_1$, where $J = 0.004|t_1|$ is the parameter of the initial exchange interaction in the PAM, which is induced by the high-energy hybridization processes in the limit $U \to \infty$. Therefore, the N\'{e}el temperature increases. Note that the effective exchange parameters $J_{\text{eff}}$ agrees well with the parameter estimated in~\cite{Das-14}.

The N\'{e}el temperature can be estimated using a simple formula similar to the formula for a quasi-2D Heisenberg antiferromagnet~\cite{Val'kov-11}:
\begin{equation}
\label{def_TN_HF}
T_\text{N} = \frac{n_{\text{L}}\pi J_{\text{eff}}}{\ln(J_{\text{eff}}/K) + c},
\end{equation}
where $c = 3.15$ and $K$ is the parameter of exchange between the nearest ions along the $z$ axis. Thus, the main effects related to the low-energy hybridization interaction in the linear portion of the dependence $T_{\text{N}}(P)$ can be approximately reduced to renormalization of the parameter of exchange between the nearest Ce ions in the $(xy)$ plane. As the pressure is increased, $n_{\text{L}}$ decreases and parameter $\tilde{A}_1$ decreases as well. However, at approaching the critical pressure, the effective exchange between more distant ions, which was not taken into account in~\eqref{def_TN_HF}, becomes important. Near the critical pressure, the frustrations arise, which are caused by the competition between the AFM and ferromagnetic exchange. As a result, the long-range order disappears.

The described effects induced by hybridization of itinerant and localized electrons in heavy-fermion antiferromagnets will take place also in 3D compounds, e.g., in CeIn$_3$. However, in the description of these compounds, it is impossible to limit the consideration to the account for hybridization and hoppings only in the $(xy)$ plane. Another interesting feature of cerium intermetallic compounds is the coexistence of superconductivity and antiferromagnetism near the quantum critical point. In the proposed model, the exchange interaction between 4f electrons can induce the Cooper instability~\cite{Val'kov-12}. Then, the formation of superconductivity near the quantum critical point can be unrelated to quantum fluctuations and be explained by the fact that antiferromagnetism suppresses Cooper pairing. However, these problems lie beyond this study.

\section{Conclusions}

Using the periodic Anderson model, we investigated the interference of two microscopic mechanisms of the formation of exchange interaction between f electrons in the quasi-2D heavy-fermion cerium intermetallic compounds.

The first mechanism is implemented at a large value of the intraatomic Coulomb repulsion and is caused by the high-energy hybridization between itinerant p electrons and f electrons of rare-earth ions. The intensity of exchange coupling determined by this mechanism is independent of temperature, concentration of itinerant carriers, and position of the chemical potential level.

The situation is qualitatively different for the second mechanism initiated by the low-energy hybridization between the above-mentioned electron groups. The contribution of these processes in the resulting exchange coupling between Ce ions significantly depends on the Fermi level position and density of states of itinerant electrons. This conclusion follows from the analysis of the behavior of magnetization of the antiferromagnetic sublattice obtained with regard to the contributions of these processes and calculated using a diagram technique for Hubbard operators. It was demonstrated that the account for the second mechanism plays a decisive role in the satisfactory description of the experimental data obtained for the quasi-2D cerium systems, e.g. CeRhIn$_5$.

\section{Acknowledgments}

This study was funded by RFBR in part according to the research projects Nos. 13-02-00523-a and 15-42-04372-r-sibir'-a. A.O.Z. is grateful for support of the Grant of the President of the Russian Federation SP-1370.2015.5.

\bibliographystyle{aipnum4-1} 
\bibliography{Valkov_arXive_2015} 

\end{document}